\documentclass[conference]{IEEEtran}
\IEEEoverridecommandlockouts
\usepackage{cite}
\usepackage{amsmath,amssymb,amsfonts}
\usepackage{graphicx}
\usepackage{textcomp}
\usepackage{xcolor}
\usepackage{svg}
\usepackage{siunitx}
\usepackage{xinttools}
\usepackage{tikz}
\usetikzlibrary{positioning}
\usepackage{mathtools}
\usepackage{neuralnetwork}
\usepackage{multicol}
\usepackage{multirow}
\newcommand\numberthis{\addtocounter{equation}{1}\tag{\theequation}}

\makeatletter
\def\ps@IEEEtitlepagestyle{%
\def\@oddfoot{\mycopyrightnotice}%
\def\@evenfoot{}%
}

\def\mycopyrightnotice{{\footnotesize
\begin{minipage}{\textwidth}\ \\[12pt] \centering
  978-1-6654-4389-0/21/\$31.00~\copyright~2021 IEEE. Personal use of this material is permitted. Permission from IEEE must be 
obtained for all other uses, in any current or future media, including 
reprinting/republishing this material for advertising or promotional purposes, creating new 
collective works, for resale or redistribution to servers or lists, or reuse of any copyrighted 
component of this work in other works. 
\end{minipage}}
\gdef\mycopyrightnotice{}}

\def\BibTeX{{\rm B\kern-.05em{\sc i\kern-.025em b}\kern-.08em
    T\kern-.1667em\lower.7ex\hbox{E}\kern-.125emX}}
\begin{document}

\title{Neural Predictive Control for the Optimization of Smart Grid Flexibility Schedules}

\author{\IEEEauthorblockN{Steven de Jongh}
\IEEEauthorblockA{\textit{Karlsruhe Institute of Technology (KIT)}\\
Karlsruhe, Germany \\
steven.dejongh@kit.edu}
\and
\IEEEauthorblockN{Sina Steinle}
\IEEEauthorblockA{\textit{Karlsruhe Institute of Technology (KIT)}\\
Karlsruhe, Germany \\
sina.steinle@kit.edu}
\and
\IEEEauthorblockN{Anna Hlawatsch}
\IEEEauthorblockA{\textit{elena international GmbH}\\
Berlin, Germany \\
anna.hlawatsch@elena-international.com}
\and
\IEEEauthorblockN{Felicitas Mueller}
\IEEEauthorblockA{\textit{Karlsruhe Institute of Technology (KIT)}\\
Karlsruhe, Germany \\
felicitas.mueller@kit.edu}
\and
\IEEEauthorblockN{Michael Suriyah}
\IEEEauthorblockA{\textit{Karlsruhe Institute of Technology (KIT)} \\
Karlsruhe, Germany \\
michael.suriyah@kit.edu}
\and
\IEEEauthorblockN{Thomas Leibfried}
\IEEEauthorblockA{\textit{Karlsruhe Institute of Technology (KIT)} \\
Karlsruhe, Germany \\
thomas.leibfried@kit.edu}}

\maketitle

\begin{abstract}
Model predictive control (MPC) is a method to formulate the optimal scheduling problem for grid flexibilities in a mathematical manner. The resulting time-constrained optimization problem can be re-solved in each optimization time step using classical optimization methods such as Second Order Cone Programming (SOCP) or Interior Point Methods (IPOPT).
When applying MPC in a rolling horizon scheme, the impact of uncertainty in forecasts on the optimal schedule is reduced. While MPC methods promise accurate results for time-constrained grid optimization they are inherently limited by the calculation time needed for large and complex power system models. Learning the optimal control behaviour using function approximation offers the possibility to determine near-optimal control actions with short calculation time. A Neural Predictive Control (NPC) scheme is proposed to learn optimal control policies for linear and non-linear power systems through imitation. It is demonstrated that this procedure can find near-optimal solutions, while reducing the calculation time by an order of magnitude. The learned controllers are validated using a benchmark smart grid.
\end{abstract}


\section{Introduction}
Future power grids are facing a variety of challenges that need to be handled to ensure a reliable and cost-efficient provision of electrical energy. On the generation side, the continuous addition of more distributed generators leads to volatile generation profiles. Additional uncertainty is added on the consumption side with the inclusion of new consumers, such as electric vehicles. Both changes require a rethinking of the operation of electrical power grids and a restructuring from a top-down to a bottom-up system. While conventional grid expansion measures can be used to make this new system resilient, these changes entail large financial costs and are not feasible in many places without substantial intervention in the daily lives of many individuals. \newline
Smart grid technologies in contrast try to solve problems of volatile generation and demand by utilizing intelligent algorithms and flexibilities in the systems. These flexibilities may include energy storage systems, heat pumps, curtailable generation and power to gas, as well as gas to power technologies. By making use of the temporal degree of freedom, the grid operator is able to choose when these grid elements operate and thus can purposefully reduce the loading on grid elements and reduce operating costs. By using models of the electrical components, grid topology and forecasts of consumption and generation, an optimal utilisation schedule of flexibilities in the grid is calculated using MPC. The goal is to find set points of flexible grid devices that minimize a user defined objective function. In most cases this objective function models the cost of grid operation while constraints ensure that technical limits are not exceeded and physical equilibria are fulfilled.

In the literature different approaches to solving this optimization problem exist. In \cite{zimmerlinFlex}, a linear formulation of a sector-coupled distribution system is optimized by using MPC. Nonlinear system models are used in \cite{dopfsgill}. In \cite{dopfhuebner}, large transmission systems are optimized using non linear system models. Especially for large systems, these methods require high calculation times. Since the optimization time step width is directly bounded by the inference time of the applied algorithms, the search for fast optimizers is crucial. In \cite{zamzam2019fastOPF}, neural networks are used to learn the solution to static optimal power flow (OPF) problems in a supervised manner. \cite{DeepOPFscDCOPFnn} and \cite{canyasse2016supervised} learn a controller for linear DC-OPF problems. Geometric Deep Learning is used in \cite{owerko2019graphopf} to learn optimal solutions to static OPF problems. Model predictive controllers for building energy optimization are learned in \cite{drgona2020deepLearningNonlinMPC}. In \cite{AKESSONnnmpc} and \cite{hertneck2018learningmpc} theoretical foundations and guarantees on learned model predictive controllers are established. \\
In this work supervised learning methods are used to imitate the behavior of optimizers for MPC. In contrast to previous applications, where solvers for static OPF problems are learned, a dynamic problem with storage elements is considered. Hyper parameter optimization is applied using Bayesian optimization (BO). The learned neural predictive controller (NPC) is compared to the classical optimization solver performance. The performance on a benchmark distribution grid with respect to objective function value and calculation time is compared.

\section{Model Predictive Control for Flexibility Scheduling}
\subsection{Optimization Problem Formulation}
The optimization of flexibility schedules can be formulated as a mathematical optimization problem. By finding the optimal state vector $x^*$ containing the optimal control actions, an arbitrary cost function $J(x^*)$ is minimized over a control horizon $H$. The optimization is subject to equality \eqref{eq:constraint2} and inequality constraints \eqref{eq:constraint1} that depend on the physical model of the system. $\mathcal{H}$ is the set of all discrete optimization time steps in the horizon. $\mathcal{I}$ and $\mathcal{E}$ are the sets of inequality and equality constraint indices.

\begin{subequations}
\begin{alignat}{4}
&\!\min_{x}        &\qquad& J(x) & \quad  & \, \label{eq:optProb}\\
&\text{subject to} &      & g_{i,t}(x) \geq 0, & \quad & \forall i \in \mathcal{I} \quad \forall t \in \mathcal{H} \label{eq:constraint1}\\
&                  &      & h_{e,t}(x) = 0. & \quad &  \forall e \in \mathcal{E} \quad \forall t \in \mathcal{H} \label{eq:constraint2}\\
&                  &      & lb_t \leq x \leq ub_t & \quad & \forall t \in \mathcal{H} \label{eq:constraint3}
\end{alignat}
\end{subequations}
The functions $g$, $h$ and $J$ can be arbitrary functions that can either be non-linear or linear. The properties of these functions determine the complexity of the underlying optimization problem and the methods that can be used for optimization.

\subsection{Electrical System Model}
In the case of smart grids, the system can be described by the steady state load flow equations. Since these equations are non-linear and non-convex they can only be solved with nonlinear programming (NLP) techniques. Since these methods require high calculation times,  linearized versions are often used. In the case where $g$, $h$ and $J$ are linear functions, this allows the application of linear programming (LP) techniques. In this paper, the manifold method from \cite{doerflerManifold} is applied to obtain linearized versions of the loadflow equations. Hence, the equality constraint \eqref{eq:constraint2} can be expressed as a linear matrix multiplication $A\cdot x = b$.

\begin{equation}
    \begin{bmatrix} B & 0 & 0 \\ C & D & 0 \\ 0 & E_p & E_E \end{bmatrix} \cdot \begin{bmatrix} x_{\text{grid}} \\ x_{\text{flex}} \\ x_{\text{sto}}  \end{bmatrix} = \begin{bmatrix} 0 \\ c \\ c_{\text{sto}} \end{bmatrix}
    \label{eq:axb}
\end{equation}
where $x=[x_{\text{grid}}, x_{\text{flex}}, x_{\text{sto}}]^T$ is the decision vector around a linearization point $\hat{x}=[\hat{x}_{\text{grid}}, \hat{x}_{\text{flex}}, \hat{x}_{\text{sto}}]^T$ and $c$ are the fixed node properties, such as PQ and voltage set points, linearized around $\hat{c}$. Equation \eqref{eq:xaufbau} shows the structure of the decision vector $x$. $x_{grid}$ consists of sub vectors for nodal voltages $v$, nodal voltage angles $\theta$, nodal active power injections $p$ and nodal reactive power injections $q$ for each time step and node in the system. $x_{flex}$ has two entries for each flexibility (e.g. storage) for each time step. The storage charging state of each storage element for each horizon time step is stored in $x_{sto}$.

\begin{equation}\label{eq:xaufbau}
x_{grid} = \begin{bmatrix} v(t_0) \\ \theta(t_0) \\ p(t_0) \\ q(t_0) \\ \vdots \\ v(t_H) \\ \theta(t_H) \\ p(t_H) \\ q(t_H) \end{bmatrix},\qquad
    \begin{aligned}
    &x_{flex} &= \begin{bmatrix} p_{flex}(t_0) \\ q_{flex}(t_0) \\ \vdots \\ p_{flex}(t_H) \\ q_{flex}(t_H) \end{bmatrix}\\
    &x_{sto} &= \begin{bmatrix} E_{sto}(t_{-1}) \\ E_{sto}(t_0) \\ \vdots \\ E_{sto}(t_H) \end{bmatrix}
    \end{aligned}
\end{equation}

The flat start linearization with $v=1$, $\theta=0$ and $p_{\text{flex}}=0$, $q_{\text{flex}}=0$ at all nodes is used. In equation \eqref{eq:axb}, the first row ensures that a valid solution of the powerflow equations is found. The second row ensures conservation of energy. In the third row the temporal coupling of storage energy is ensured. The entries of $B$, $C$ and $D$ are obtained as block diagonal matrices $B=\text{diag}(\Tilde{B})$, $C=\text{diag}(\Tilde{C})$ and $D=\text{diag}(\Tilde{D})$ where the entries $\Tilde{B}, \Tilde{C}, \Tilde{D}$ are obtained using the manifold method \cite{doerflerManifold}.  $E_p$ and $E_E$ are matrices that are used for temporal coupling of the storage equations. $E_p$ is used to calculate the energy of storages based on the their flexible power injection. $E_E$ simulates self discharging of the batteries. The resulting matrices do not change for the optimization horizon time steps. Equation \eqref{eq:batteryequ} shows the underlying equations for the storage system under the assumptions of a charging and discharging efficiency of 100 \%.

\begin{align*}
    E_{sto,t+1} &= E_{sto,t} \cdot (1-\mu_{sd} \cdot \Delta t) + p_{sto,t} \cdot \Delta t \\
    -1 &\leq \frac{p_{sto,t}}{p_{sto,max}} \leq 1 \numberthis \label{eq:batteryequ} \\ 
    0 &\leq \frac{E_{sto,t}}{E_{sto,max}} \leq 1
\end{align*}

Equation \eqref{eq:boundscurrents} shows the linearization of bounds on the branch currents which is used to limit the flows on the branch elements. The inequality constraints in \eqref{eq:constraint1} can be expressed in linear form as shown in equation \eqref{eq:gxh}. Here, $G_l$ are incidence matrices and $I$ are identity matrices.

\begin{equation}
    \begin{aligned}[b]
    & |v_i - v_j| = \\
    &\begin{cases}
    v_i - v_j \leq (\hat{v}_j - \hat{v}_i) + \frac{I_{\text{eff,max,ij}} \cdot \sqrt{3} \cdot V_b}{|y_{ij}|\cdot S_b}, \quad \text{if} \, v_i \geq v_j \\
    v_j - v_i \leq (\hat{v}_i - \hat{v}_j) + \frac{I_{\text{eff,max,ij}} \cdot \sqrt{3} \cdot V_b}{|y_{ij}|\cdot S_b}, \quad \text{if} \, v_i \leq v_j
    \end{cases}
    \end{aligned}
    \label{eq:boundscurrents}
\end{equation}

\begin{equation}
\begin{bmatrix}
G_l & 0 & 0 \\
-G_l & 0 & 0 \\
I_v & 0 & 0 \\
0 & I & 0 \\
0 & 0 & I \\
-I_v & 0 & 0 \\
0 & -I & 0 \\
0 & 0 & -I
\end{bmatrix} \cdot
\begin{bmatrix}
x_{\text{grid}} \\ x_{\text{flex}} \\ x_{\text{sto}} 
\end{bmatrix}
\leq
\begin{bmatrix}
ub_{\text{lines}} \\ lb_{\text{lines}} \\ ub_{x_{\text{grid}}} \\ ub_{x_{\text{flex}}} \\ ub_{x_{\text{sto}}} \\ lb_{x_{\text{grid}}} \\ lb_{x_{\text{flex}}} \\ lb_{x_\text{sto}}
\end{bmatrix}
\label{eq:gxh}
\end{equation}

A quadratic objective function shown in equation \eqref{eq:objective} is chosen.
\begin{equation}
    J(x) = \frac{1}{2} \cdot x^T \cdot M \cdot x + d \cdot x
    \label{eq:objective}
\end{equation}
In \eqref{eq:objective}, $M$ is a binary matrix that gathers the slack power $p_{\text{slack}}$ at all time steps. $d$ is used to substitute for the offset due to linearization. Since the slack node is chosen as the connection to the external grid this leads to a minimization of the exchange of power with the higher voltage level and maximizes self-sufficiency. The quadratic term in the objective function reduces peaks in the exchanged power with the higher voltage level. The overall optimization problem is a linear constrained quadratic program (LCQP) and can be solved with appropriate optimiation algorithms. Since the underlying system is a system consisting of sparse matrices, due to the sparsity of the electrical system and temporal structure, MOSEK \cite{mosek} is used as a suitable solver. Calculations are performed on a \textit{Ryzen Threadripper 3990X} CPU. A time discretization of $\Delta t = 0.25 \, h$ and time horizon $H=24 \,h$ is chosen. Furthermore, the voltage bounds are set to $v_{ub}=1.05 \, p.u.$ and $v_{lb}=0.95 \, p.u.$ for all system buses.

\section{Neural Predictive Control}

\subsection{Neural Network Imitation Learning}
The goal of the NPC is to imitate the behaviour of the MPC from data generated by the MPC. For this purpose, the NPC uses a learnable function approximator to learn the function that maps the system state to optimal control outputs. A neural network is used as a function approximator due to its flexible training and its universal approximation properties. The general equation of a three layer feed forward neural network for regression tasks is shown in equation \eqref{eq:neuralnetwork}. The neural network consists of learnable parameters $W_i$ and $b_i$, where $i$ denotes the layer the weight matrix and biases are in. $\sigma_i$ are the activation functions of the hidden layers. The last layer has no activation function to allow unbounded outputs. In this paper, pytorch \cite{pytorch} is used for the implementation of neural networks. All inference and training is performed on a \textit{TITAN RTX} graphics card.

\begin{equation}
    y = W_3\cdot (\sigma_2(W_2 \cdot (\sigma_1(W_1 \cdot x + b_1 )) + b_2)) + b_3
    \label{eq:neuralnetwork}
\end{equation}

\begin{figure}[htbp]
    \centering
    \includegraphics[width=\columnwidth]{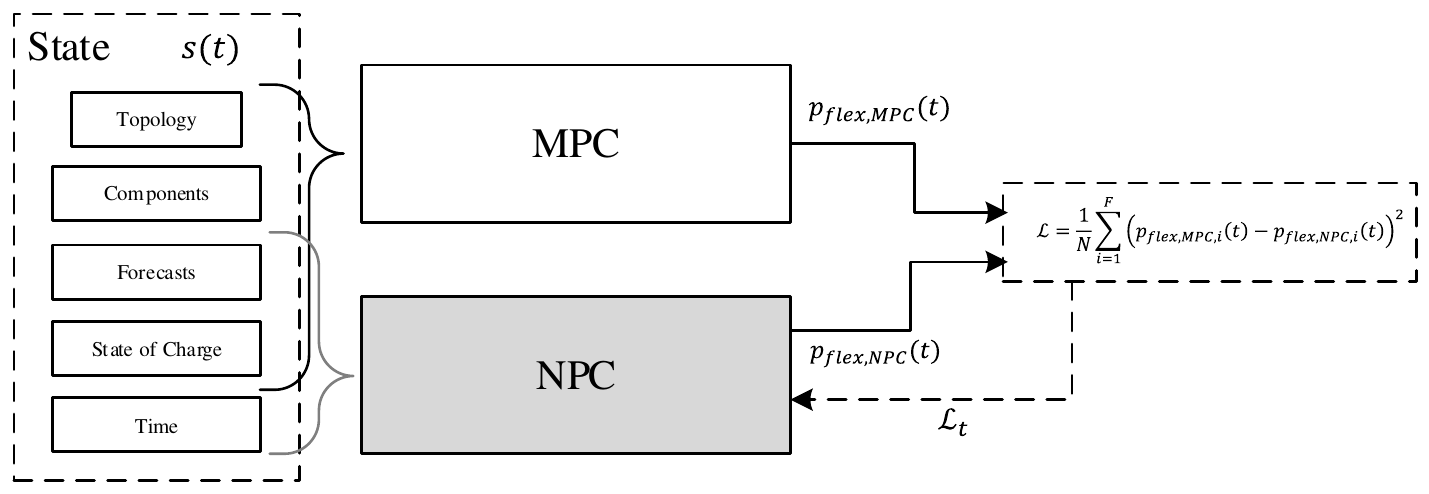}
    \caption{Imitation learning process}
    \label{fig:imitationprocess}
\end{figure}

Fig. \ref{fig:imitationprocess} shows a schematic of the imitation learning process. The goal is to learn the behaviour of the MPC algorithm on the grid data. For this purpose each grid state is fed into the MPC algorithm, as well as the NPC. The MPC takes the topology, as well as position of grid elements such as storages, generators etc. as input. Furthermore, the forecast of consumption and generation for the next day is taken into account. Given the current state of charge of the storages, the MPC calculates the optimized schedule $p_{flex,MPC}(t)$ for all flexible devices for the forecast horizon. Since the topology and component positions are assumed to be fixed over the simulation time the NPC only takes the forecasts and current battery state of charge as input. Using the feed forward operation, the NPC predicts a schedule $p_{flex,NPC}(t)$. Since this schedule is not guaranteed to be optimal a loss $\mathcal{L}_t$ is calculated based on the mean squared error (MSE) between the MPC and NPC output. This loss is used to calculate gradients on the weights $W_i$ and biases $b_i$ of the neural network, allowing the neural network to learn the behaviour of the MPC over multiple iterations.

Since the training procedure of NPC in fig. \ref{fig:imitationprocess} can be performed independently of another, the following steps are performed:
\begin{enumerate}
    \item \textbf{Rolling horizon MPC}: The MPC is performed for a year using a rolling horizon approach, given load and generation timeseries. The input data of the MPC is saved as tensor $X$ and the outputs are saved as tensor $Y$
    \item \textbf{Data Splitting and Scaling}: To make the performance of both algorithms comparable, the data is split into a training and testing data set. Furthermore, the data is rescaled to improve the training of the neural network.
    \item \textbf{Training of NPC}: The neural network of the NPC is trained with the training data generated by step 2).
\end{enumerate}
The first step applies the MPC to each timestep $\Delta t=0.25\,h$ of a year. The given forecasts for the horizon of 24 hours are used for optimization of the schedule of flexible power for the next timestep $\Delta t$. The rolling horizon applies this procedure at each time step. The forecasts are assumed to be without error. In real applications, where forecast errors occur, the rolling horizon scheme has the goal to re-calculate the optimal schedule regularly to use the most recent forecasts and take the new time step $H+1$ into account. The input and output data of the MPC are used as training data. The obtained data set $(X, Y)$ is split into a training $(X_{\text{train}}, Y_{\text{train}})$ and testing data set $(X_{\text{test}}, Y_{\text{test}})$ using a 75 \% / 25 \% split. To ensure that the training data is chosen equally from all seasons and to maintain temporally coherent periods of time, three weeks of each month are used as training data. The remaining week of each month is used for testing.

\section{Benchmark Distribution Grid}
\begin{figure}[htbp]
    \centering
    \includegraphics[width=\columnwidth]{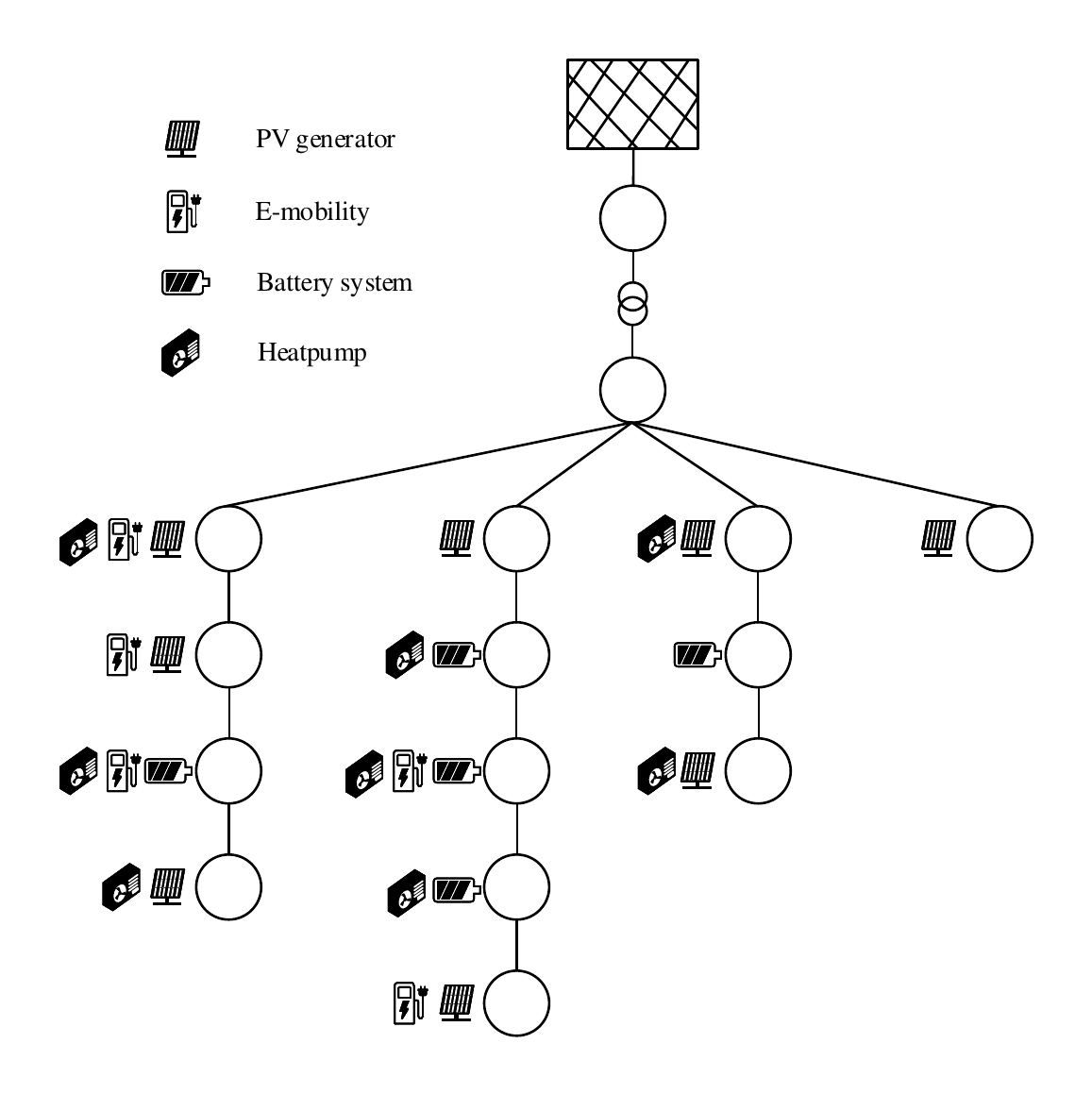}
    \caption{Schematic of 15 node benchmark grid (scenario 2034) \cite{meinecke2020simbench}}
    \label{fig:benchmarkgrid}
\end{figure}

To test the performance of the MPC and NPC algorithms, a benchmark distribution grid is used. The grid is based on the \textit{LV1} network of simbench \cite{meinecke2020simbench}. The grid consists of 15 nodes and 13 households and has radial structure. The grid is intended to simulate a future scenario for 2034 with a high penetration of electric vehicles and sector coupling.
A schematic of the benchmark grid is shown in fig. \ref{fig:benchmarkgrid}. 
The available time series allow a simulation of one year with pandapower \cite{thurner2018pandapower} using time steps of $\Delta t = 0.25 \, h$. Due to the high penetration of sector coupling and electric vehicle charging infrastructure, the transformer which connects the low voltage grid with the medium voltage grid is overloaded in several time steps.

\section{Results}

For hyperparameter optimization Bayesian optimization is applied for 180 iterations. In each iteration BO trains a feed forward neural network with hyperparameters chosen in a given range. Table \ref{tab:hyperopt} shows the bounds and results of the hyperparameter optimization. Note that these hyperparameters are specific for the used benchmark grid and can be different for other grids. However, hidden layers with 1.5 times the input dimension showed the best results on multiple benchmarks. The optimized neural network consists of three hidden layers. For the given setup the input has 2792 elements. The output has 480 elements. Rectified linear units are chosen as activation functions for all but the last layer.

\begin{figure}[htbp]
    \centering
    \includegraphics[width=\columnwidth]{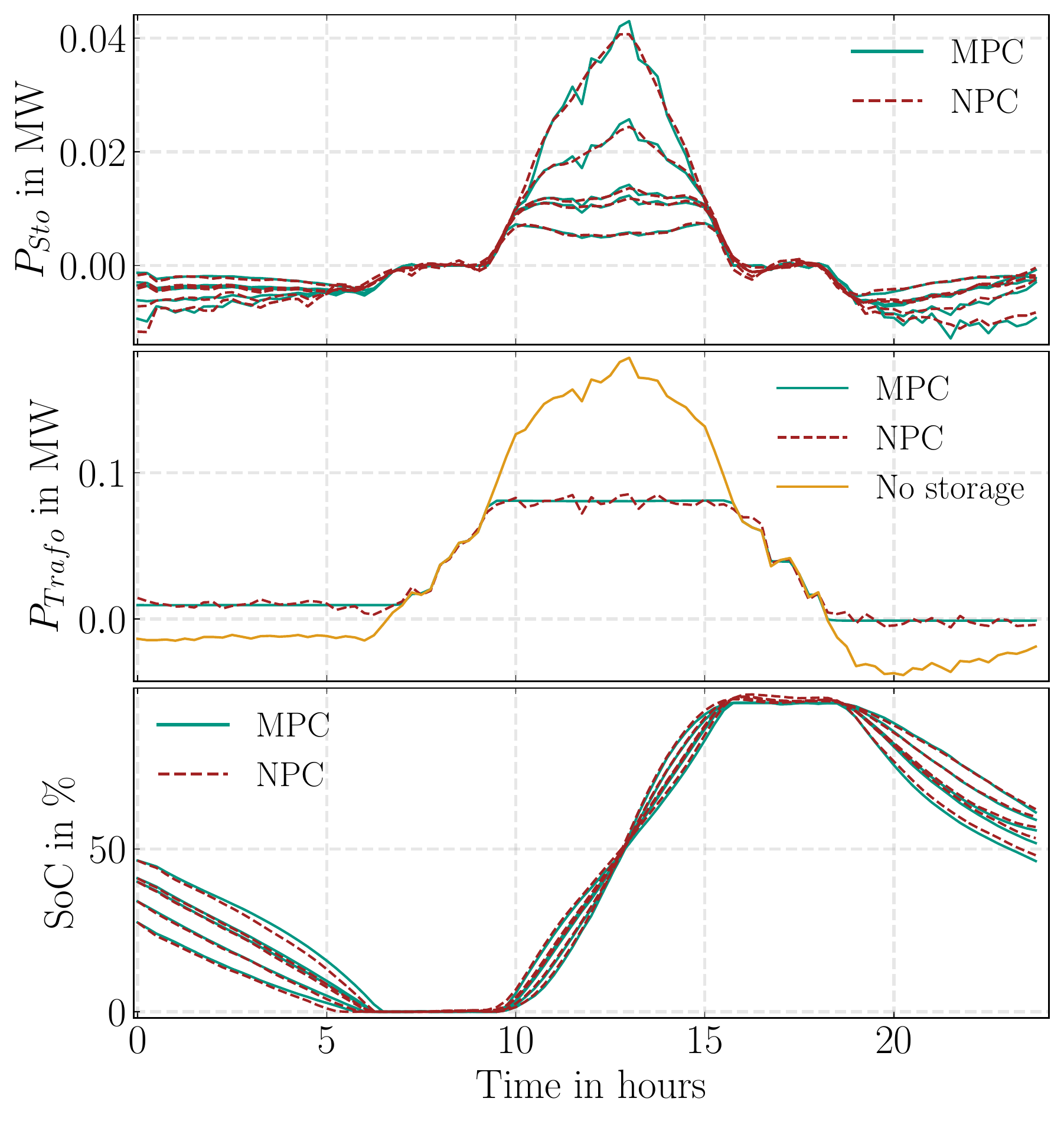}
    \caption{Optimization of MPC and NPC for single day}
    \label{fig:einzelzeitpunkt}
\end{figure}

\begin{table}[htbp]
    \caption{Results of hyperparameter optimization using BO}
    \begin{center}
    \begin{tabular}{|c||c|c|c|}
    \hline
    \textbf{Parameter} & \textbf{Lower bound} & \textbf{Upper bound} & \textbf{Optimized value} \\
    \hline
    learning rate $\alpha$ & $10^{-6}$ & $10^{-3}$ & $6.752 \cdot 10^{-4}$ \\
    \hline
    epochs & $15$ & $500$ & $500$ \\
    \hline
    hidden layers & $1$ & $4$ & $3$ \\
    \hline
    neurons $L_1$ & $240$ & $5584$ & $4323$ \\
    \hline
    neurons $L_2$ & $240$ & $5584$ & $2565$ \\
    \hline
    neurons $L_3$ & $240$ & $5584$ & $2194$ \\
    \hline
    \end{tabular}
    \label{tab:hyperopt}
    \end{center}
\end{table}
The neural network with the optimized hyperparameters is trained given the data set of the benchmark system that is constructed using the previously described procedure. The MPC, as well as the NPC are applied in a rolling horizon fashion with the task to minimize the exchange with the external grid and constrained by bus voltages and branch flows limits. The five battery storage systems shown in fig. \ref{fig:benchmarkgrid} are used as control output of both algorithms.
Fig. \ref{fig:einzelzeitpunkt} shows the results of both MPC and NPC, as well as the baseline of no storage usage on a single day. For the given fig., the 11th of june was chosen, starting at midnight. The first plot shows the controlled storage power of all five storages in the system. It can be seen, that the storages are discharged at night (hours 0 to 7 and 19 to 24) and charged during the day (hours 9 to 16). This results in a peak shaving of the transformer power as can be seen in the second subplot. This is an expected outcome, as the objective function aims at minimizing the exchange of power with the medium voltage grid. Both algorithms use the storage capacity between 0 and 80 \%. This is due to the maximum storage SoC that is set at 80 \%. As can be seen, the NPC accurately approximates the storage powers for all five storages. Overall the output of the NPC has a smoother curve. However, this leads to fluctuations in the transformer power, which is constant for the MPC approach since fluctuations in generation and load are smoothened in a mathematical exact way.

\begin{figure}[htbp]
    \centering
    \includegraphics[width=\columnwidth]{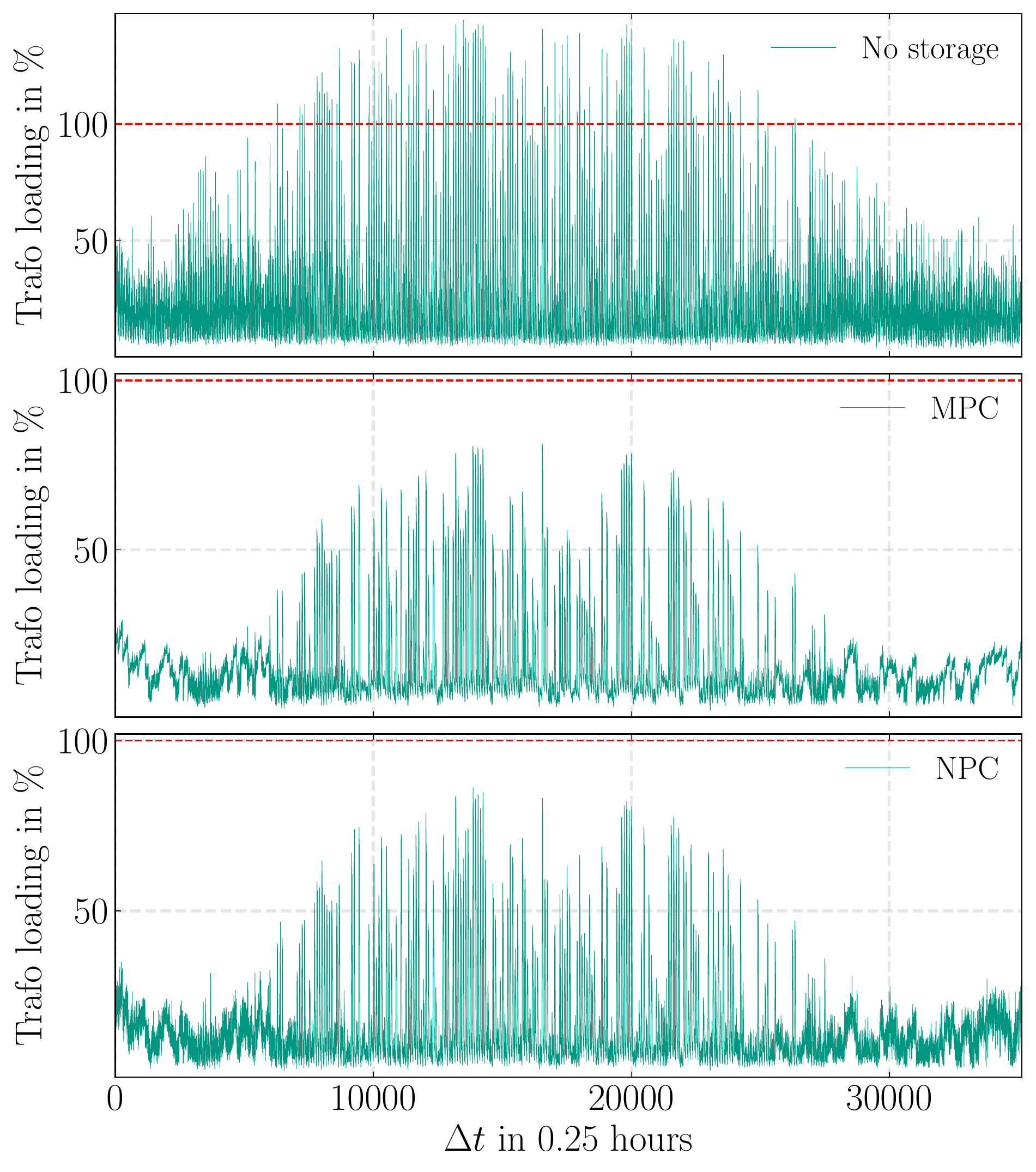}
    \caption{Grid loading \textit{LV1} for yearly simulation}
    \label{fig:auslastungsplot}
\end{figure}

Fig. \ref{fig:auslastungsplot} shows the transformer loading over the yearly simulation. The objective of both the MPC and NPC is to minimize this loading. As can be seen, when not using the installed storage system the transformer is overloaded during many time steps, especially in summer, where PV generation is high. Both MPC and NPC succeed in reducing the transformer loading below the maximum allowed transformer power for all time steps of the year. The effect of the NPC is similar to the MPC while showing slightly higher fluctuations in the transformer loading.

\begin{figure}[htbp]
    \centering
    \includegraphics[width=\columnwidth]{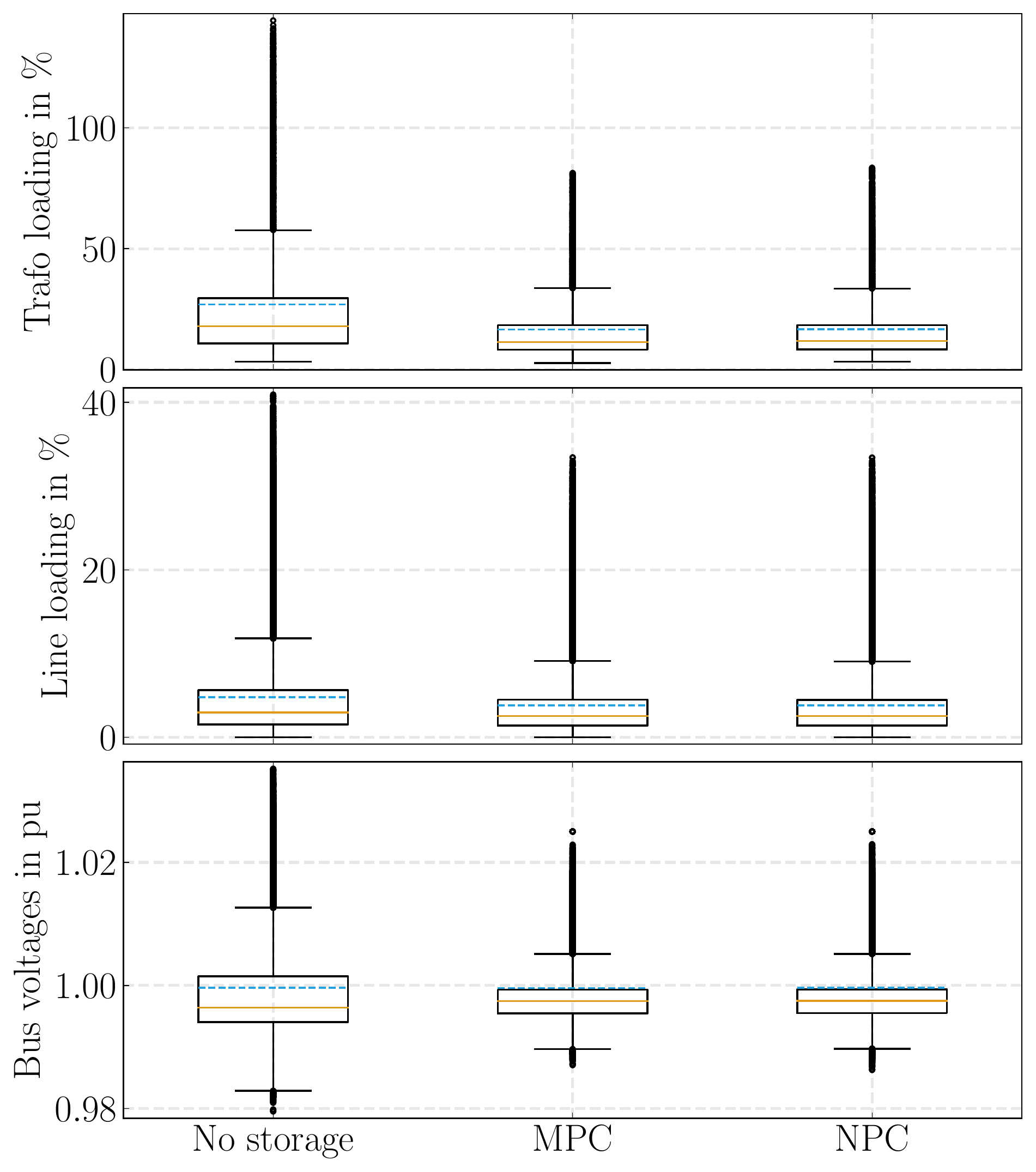}
    \caption{Constraint violations over data set for \textit{LV1}}
    \label{fig:constraintviolations}
\end{figure}

Since the controllers should ensure that the constraints of transformer loading, line loading and bus voltages are not violated, fig. \ref{fig:constraintviolations} compares the statistical properties of these metrics over the yearly simulation of \textit{LV1}. It can be seen, that the transformer loading exceeds the maximum allowed value of 100 \% during multiple time steps in the year. MPC as well as NPC succeed in reducing the maximum loading well below 100 \%. Furthermore, both MPC and NPC have similar behavior with respect to their mean, median and outliers. It can be seen that NPC successfully learned the behavior of MPC and is able to reproduce the behavior with respect to transformer constraints. Similar behavior can be seen for the line loading. Again NPC is able to reproduce the results from MPC. In this case the base case without battery usage did not violate the line loading constraints during the whole year. The voltages are limited to $\pm 5\,\%$ of the nominal voltage (400 Volt), leading to an allowed range of $0.95$ to $1.05$ $p.u.$. The voltage limits are not violated in the given data set. Nevertheless, the operation of the battery system leads to more uniform voltages and significantly reduces the covered voltage range for both MPC and NPC. Again the statistical effect on the results of both algorithms is similar.

\begin{figure}[htbp]
    \centering
    \includegraphics[width=\columnwidth]{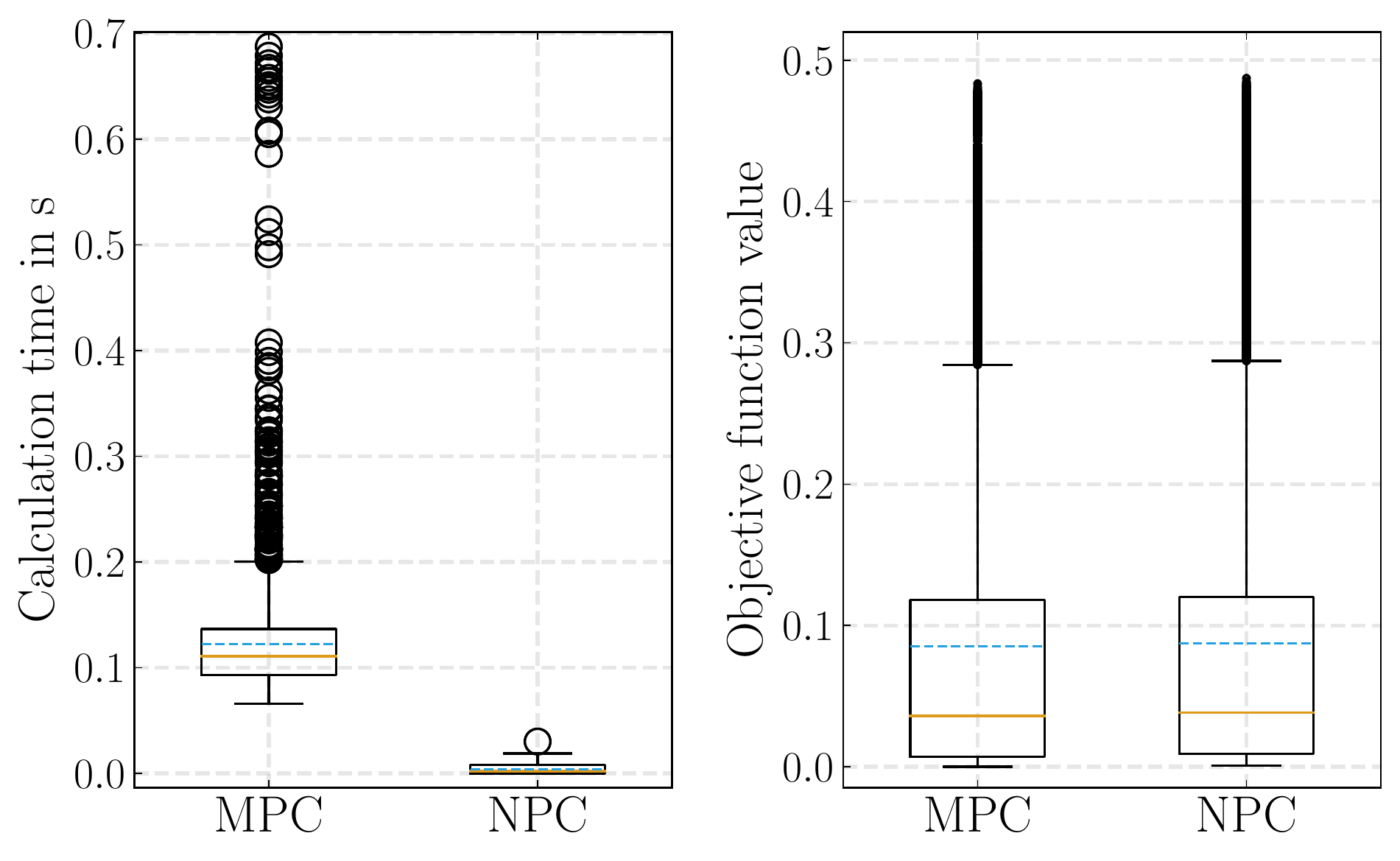}
    \caption{Calculation time and objective function value for year simulation \textit{LV1}}
    \label{fig:rechenzeit}
\end{figure}

For the given example grid the objective function aims at minimizing the exchanged power with the medium voltage grid (e.g. minimizing the power flow over the transformer). Both MPC and NPC can be compared with respect to their achieved objective function values. Fig. \ref{fig:rechenzeit} compares the calculation time and objective function value of both MPC and NPC on the data set of \textit{LV1}. The objective function value distribution has similar statistical properties for both algorithms, again highlighting the successful imitation of the behaviour of the MPC by the NPC. It can be seen that the calculation time can be reduced by an order of magnitude when using NPC for control. Additionally, MPC is subject to outliers in calculation time that can be up to seven times the median value. For NPC all but one time step are within close range of the median value.

\begin{table}[htbp]
    \centering
    \caption{objective statistics over year, $\textit{LV1}$}
    \begin{tabular}{|c|c|c||c|}
        \hline
        & \textbf{MPC} & \textbf{NPC} & \textbf{Ratio} \\
        \hline
        objective mean & 0.0786 & 0.0797 & 1.0139 \\
        \hline
        objective median & 0.0318 & 0.0339 & 1.0684 \\
        \hline
        inference time mean & 259.0 ms & 4.7 ms & 0.0181\\
        \hline
    \end{tabular}
    \label{tab:objectivevals}
\end{table}

The performance statistics are summarized in table \ref{tab:objectivevals}. On the overall data set of \textit{LV1} the objective function value of NPC is $1.39 \, \%$ higher compared to MPC. The median increases by approximately $6.84 \, \%$. Both results highlight that NPC is able to learn the behaviour of MPC in a precise manner. However, small decreases in performance are to be expected when learning MPC through imitation. For the given grid the mean inference time for the MPC is 259.0 ms, while the mean inference time of the NPC is 4.7 ms. This is an increase of approximately a factor of 55.


\section{Conclusion \& Outlook}
In this work, the imitation of MPC through fast function approximators for the task of smart grid flexibility scheduling is undertaken. It is shown, that the performance of NPC is in the range of classical MPC methods when learning the optimal controllers through imitation. While NPC has slightly higher objective function values, it can improve the control task by offering a calculation time reduction by one order of magnitude. Tests on a benchmark smart grid show, that the NPC successfully removed all constraint violations on power flows and bus voltages. While this shows promising results for a single benchmark system, more extensive studies on the constraint violation properties are needed to make general conclusions. Especially for systems that operate close to their limits (e.g. constraints), conclusions are of interest. The proposed imitation learning scheme learns an MPC controller that is based on LCQP. Under the assumption of the non-linear loadflow equations these results can be seen as an approximation with respect to the underlying physical system but as the best case with respect to calculation time. Hence, for the non linear programming (NLP) higher calculation times are to be expected. A comparison of learned controllers and NLP is an important area of research. The proposed NPC controller learns the optimal behaviour for a given grid configuration and MPC algorithm. To gain inductive behaviour that can be transferred to arbitrary grid topologies, research in the direction of geometric deep learning techniques is of interest. Additionally, usage of recurrent structures that learn on the temporal structure of the problem are a possible approach. By making use of function approximations that are limited to convex functions, f.e. using input-convex neural networks, faster controllers with guarantees can be derived.

\bibliographystyle{IEEEtran}
\bibliography{bibliography.bib}

\end{document}